\documentclass[prc,twocolumn]{revtex4}

\usepackage{graphicx}
\usepackage{dcolumn}
\usepackage{bm}
\usepackage{amsmath}

\begin{document}

\title[Electron-capture decay in transfermium isotopes]
{Electron-capture decay in isotopic transfermium chains from 
self-consistent calculations}

\author{P.~Sarriguren}
\affiliation{Instituto de Estructura de la Materia, IEM-CSIC, Serrano 123, 
E-28006 Madrid, Spain}
\email{p.sarriguren@csic.es}
\vspace{10pt}

%\date{\today}

\begin{abstract}
Weak decays in heavy nuclei with charge numbers $Z=101-109$ are studied within 
a microscopic formalism based on deformed self-consistent Skyrme Hartree-Fock 
mean-field calculations with pairing correlations. The half-lives of $\beta^+$ 
decay and electron capture are compared with $\alpha$-decay half-lives obtained
from phenomenological formulas. Transfermium isotopes of Md, No, Lr, Rf, Db, Sg, 
Bh, Hs, and Mt that can be produced in the frontier of cold and hot 
fusion-evaporation channels are considered. Several isotopes are identified whose 
$\beta^+/EC$- and $\alpha$-decay half-lives are comparable. The competition between 
these decay modes opens the possibility of new pathways towards the islands of
stability.

\end{abstract}

%\noindent{\it Keywords}: Weak-decay half-lives, superheavy nuclei, nuclear 
%energy-density functional
%\maketitle

%%%%%%%%%%%%%%%%%%%%%%\maketitle
\maketitle

\section{Introduction}

Understanding superheavy nuclei (SHN) is a topical issue that attracts a great deal of
research activity. The synthesis of new SHN is a long story that has already
led in the last decades to the discovery of a large number of new elements (see
\cite{hofmann_00,oganessian_07,hamilton_13,oganessian_15,oganessian_15_npa,hofmann_16,giuliani_19} for a review).

Cold-fusion reactions were first used to synthesize SHN with $Z=107-113$ by using
target nuclei $^{208}$Pb and $^{209}$Bi and medium-mass stable isotopes of Ti, Cr, Fe,
Ni, and Zn as projectiles \cite{hofmann_00,hamilton_13,morita_04}.
In these cold reactions the compound nucleus produced has relatively low excitation 
energy and typically evaporates one or two neutrons. Attempts to produce heavier 
nuclei with these type of reactions failed because of the fast decreasing of the
production cross sections for increasing charge of the projectiles. This difficulty
was solved by using more asymmetric reactions with both target and projectile having 
a large neutron excess and thus, decreasing the Coulomb repulsion. 
In practice, these hot-fusion reactions were carried out with long-lived actinide 
nuclei from $^{238}$U to$^{249}$Cf as targets and the double magic nucleus $^{48}$Ca 
as projectiles. The method was successfully accomplished to produce 
SHN with $Z$=112--118 in the neutron-evaporation channels ($xn$-channels) 
\cite{oganessian_07,oganessian_15,oganessian_04}. The compound nuclei produced in 
these hot reactions are created in highly excited states that evaporate typically 
between 2 and 5 neutrons before starting a chain of $\alpha$ decays ending with a 
spontaneous fission (SF). Tracing back these paths allows one to arrive to the SHN
originally produced.

However, the SHN synthesized so far are still far from the theoretically predicted 
"islands of stability" for SHN. Calculations of binding energies within 
macroscopic-microscopic models \cite{myers_66,sobiczewski_66,nilsson_68,patyk_91,moller_94,smolanczuk_95,kuzmina_12} predict the existence of particularly stable 
structures for spherical SHN with $Z=114$ and $N=184$, as well as for deformed 
configurations with $Z=108$ and $N=162$. These results agree with more recent 
microscopic calculations performed within self-consistent relativistic and 
non-relativistic mean-field models 
\cite{rutz_97,kruppa_00,bender_01,bender_03,meng_06,dobaczewski_15} 
that predict stabilized regions with shell closures at $Z=114,\ N=184$, $Z=120,\ N=172$, 
and $Z=126,\ N=184$, depending on the interactions and their parametrizations. 

Further experimental investigation into the more neutron-rich SHN region following
the direction of the predicted island of stability is a difficult task in the $xn$ 
channels because of the limited number of available stable projectiles and targets 
to reach those nuclei, as well as because of the small production cross sections. 
Studies of the optimal combinations of target and beam partners in a search for
more efficient reactions to synthesize SHN can be found in 
\cite{adamian_04_1,adamian_04_2,adamian_04_3}.

Alternative ways to produce more neutron-rich isotopes through fusion-evaporation 
reactions that include not only $xn$ channels, but also the emission of charged 
particles from the compound nucleus ($pxn$ and $\alpha xn$ channels), are presently 
being explored. The products of these reactions might fill the gap of unknown
nuclides between those produced in cold and hot reactions in the $xn$ 
fusion-evaporation channels.
A number of such reactions, including isotopes of Md, No, Lr, Rf, Db, Sg, Bh, Hs, 
and Mt have been studied in asymmetric hot fusion-evaporation reactions 
\cite{hong_15,hong_16}, predicting cross sections 
that are about one or two orders of magnitude smaller than those of the $xn$ channels. 
Nevertheless, although the cross sections are smaller, the $pxn$ and $\alpha xn$ 
channels will allow the production of new isotopes that are unreachable 
in the $xn$ channels due to the lack of proper projectile-target combinations.
Production of superheavy isotopes with $Z=111-117$ has also been studied in
the charged particle-evaporation channels of $^{48}$Ca-induced actinide-based
fusion reactions \cite{hong_17}.
Recent measurements \cite{lopez_19} or reanalysis \cite{hessberger_19} of 
proton-evaporation rates in the $^{50}$Ti($^{209}$Bi,$xn$)$^{259-x}$Db reactions 
have demonstrated the viability of these fusion reactions to approach the island 
of stability. The average cross sections for proton evaporation are found 
experimentally to be between 10 and 100 times smaller than the cross sections 
in the neutron-evaporation channels. Although small, the former may represent 
an alternative way to produce more neutron-rich SHN.

Similarly, $\beta^+/EC$-decay may open new pathways towards the predicted region 
of stability and may help to fill the gap between the nuclei produced in cold 
and hot fusion reactions \cite{karpov_12,zagrebaev_12}.
Experimental evidence of this decay has been already found in $^{258}$Db
\cite{hessberger_16}. However,
the $\beta^+/EC$-decay half-lives in SHN have not been sufficiently studied yet.
There are phenomenological parametrizations \cite{zhang_06} that can be used to 
extrapolate to regions where the half-lives are unknown. There are also calculations 
that neglect nuclear structure effects, such as those in 
\cite{karpov_12,zagrebaev_12}, where only allowed transitions connecting
parent and daughter ground states are considered. The nuclear matrix elements 
of the transitions were assumed to be a constant phenomenological value given 
by $\log(ft)=4.7$ for all nuclei. This could be a rather low value leading to 
short half-lives, if one compares it with other values between 5.7 and 6.5
that can be found in the literature \cite{fiset_72} within a similar approach.
$Q_{EC}$ energies were taken
from the masses of the finite range droplet model (FRDM) \cite{FRDM}. 
In a different approach, half-lives for $\beta^+/EC$-decay were also evaluated 
within a proton-neutron quasiparticle random-phase approximation (pnQRPA), 
which is based on a phenomenological folded-Yukawa single-particle Hamiltonian, 
using masses from FRDM and standard phase factors. However, only $\beta^+/EC$ 
half-lives smaller than 100 s were published \cite{moller_97}.

In this work, the $\beta^+/EC$-decay half-lives of some selected even-even and 
odd-$A$ isotopes in the region $Z=101-109$ and $N=151-168$ are studied. 
Namely, $^{259}$Md, $^{260,261}$No, $^{255,257,259,261,263,265}$Lr, $^{255-267}$Rf, 
$^{257,259,261,263,265,267,269}$Db, $^{258-271}$Sg, $^{261,263,265,267,269,271,273}$Bh, 
$^{263-275}$Hs, and $^{265,267,269,271,273,275,277}$Mt. 
We use a microscopic approach based on a deformed self-consistent Hartree-Fock 
calculation with Skyrme interactions and pairing correlations that describes 
the nuclear structure of parent and daughter nuclei involved in the 
$\beta^+/EC$-decay. The method has been already used to calculate those half-lives 
in different SHN \cite{sarri_19} and therefore, the present work is an extension
of the previous study to nuclei in the transfermium region.

The  paper is organized as follows. Section II contains the basis of the  
theoretical method used to calculate Gamow-Teller (GT) strength distributions and 
$\beta^+/EC$ half-lives. The phenomenological models used to calculate 
$\alpha$-decay half-lives are also presented. Section III contains the results
for the SHN mentioned above and they are discussed in terms of their relevance
in terms of the competition between $\beta^+/EC$- and $\alpha$-decay modes.
Finally, Section IV contains the summary and conclusions.

\section{Theoretical formalism}

A brief summary of the theoretical framework used in this work to account for
the $\beta^+/EC$-decay half-lives in SHN is presented in this section. 
The procedure used here follows closely the approach used in \cite{sarri_19}.
Further details of the formalism can be found elsewhere
 \cite{sarri1,sarri_99,sarri2,sarri_odd,moller1,hir2}

The $\beta^+/EC$-decay half-life, $T_{\beta^+/EC}$, is calculated by summing all the 
allowed GT transition strengths connecting the parent ground state with
states in the daughter nucleus with excitation energies, $E_{ex}$, lying below 
the $Q_i$ energy ($i=\beta^+,EC$), 

\begin{equation}
T_{i}^{-1}=\frac{\left( g_{A}/g_{V}\right) _{\rm eff} ^{2}}{D}
\sum_{0 < E_{ex} < Q_i}f^i\left( Z,Q_{i}-E_{ex} \right) B(GT,E_{ex}) \, ,
 \label{t12}
\end{equation}
with $D=6143$~s and $(g_A/g_V)_{\rm eff}=0.77(g_A/g_V)_{\rm free}$, where 0.77 is a 
standard quenching factor and $(g_A/g_V)_{\rm free}=-1.270$.
$T_{\beta^+/EC}$ is the joint contribution from both $\beta^+$ and $EC$ decays.
The energies $Q_i$ can 
be written in terms of the nuclear masses $M(A,Z)$ and the electron mass ($m_e$),

\begin{equation}
 Q_{EC}=Q_{\beta^+} +2m_e= M(A,Z)-M(A,Z-1)+m_e \, ,
\end{equation}
In this work these energies are taken from experiment \cite{audi_12}. They 
are shown in table \ref{table_q} together with their uncertainties. 

To get the half-lives, the GT strengths are weighted with phase-space factors 
$f^i(Z,Q_i-E_{ex})$. The two components of these factors, positron emission $f^{\beta^+}$ 
and electron capture $f^{EC}$, are computed numerically for each value of 
the energy, as explained in \cite{gove}. 

Concerning the nuclear structure involved in the calculation of the GT strength
$B(GT,E_{ex})$, a self-consistent calculation of the mean field is first performed 
by means of a deformed Hartree-Fock procedure with Skyrme interactions and pairing 
correlations in the BCS approximation with fixed gap parameters. This calculation 
provides us single-particle energies, wave functions, and occupation probabilities. 
The Skyrme interaction SLy4 \cite{chabanat} is chosen for this study because of its
proven ability to describe successfully nuclear properties throughout the entire 
nuclear chart \cite{bender08,stoitsov_1,stoitsov_2}. 
The solution of the HF equations is found 
by using the formalism developed in \cite{vautherin}, under the assumption of 
time reversal and axial symmetry. The single-particle wave functions are expanded 
in terms of the eigenstates of an axially symmetric harmonic oscillator in 
cylindrical coordinates, using 16 major shells. This basis size is sufficiently
large to achieve convergence of the HF energy. One should also take into account
that the use of an axially deformed harmonic oscillator basis, which is tuned
in terms of two parameters (oscillator length and axis ratio), accelerates the
convergence as compared to the spherical basis. 
Deformation-energy curves (DECs)
are constructed by constrained HF calculations that allow to analyze the nuclear 
binding energies as a function of the quadrupole deformation parameter $\beta_2$. 

Calculations of the GT strengths are performed subsequently for the deformed ground 
states that correspond to the absolute minima in the DEC. Nuclear deformation has 
been revealed as a key element to describe $\beta$-decay properties in many 
different mass regions  \cite{sarri1,sarri_99,sarri2,sarri_odd} and it is also 
expected to play a significant role in SHN \cite{sarri_19}. A deformed  pnQRPA 
with residual spin-isospin interactions is used to obtain the energy distribution 
of the GT strength needed to calculate the half-lives. In the case of SHN the 
coupling strengths of the residual interactions that scale with the inverse of 
the mass number are expected to be very small and their effect is neglected.
The GT strength distributions in the following figures are referred to 
the excitation energy in the daughter nucleus.

In the case of odd-$A$ nuclei the procedure is based on the blocking of the state 
that corresponds to a given spin and parity ($J^{\pi}$), using the equal filling 
approximation (EFA) to calculate its nuclear structure \cite{sarri_odd}. 
The EFA prescription is commonly used in self-consistent mean-field calculations
because it preserves time-reversal invariance with the corresponding numerical
advantages associated to this symmetry. In this approximation half of the unpaired
nucleon sits in a given orbital and the other half in the time-reversed partner.
The reliability of this approximation has been demonstrated \cite{schunck_10} by
comparing the results from EFA with those from more sophisticated approaches,
including the exact blocking procedure with time-odd mean fields fully taken into
account. It was shown \cite{schunck_10} that both procedures are strictly equivalent
when time-odd terms are neglected and that the impact of the time-odd terms is
quite small. The final conclusion was that the EFA is sufficiently precise for
most practical applications. 
The blocked state is chosen among the states in the vicinity of the Fermi level
as the state that minimizes the energy.

This model has been successfully used in the past to study different mass regions
including neutron-deficient medium-mass \cite{sarri_wp,sarri_rp_1,sarri_rp_2,sarri_rp_3} 
and heavy nuclei \cite{sarri_pb1,sarri_pb2,sarri_pb3}, neutron-rich nuclei 
\cite{sarripere1,sarripere2,sarripere3,kiss,sarri_rare}, and $fp$-shell nuclei 
\cite{sarri_fp1,sarri_fp2_1,sarri_fp2_2}.
The effect of various ingredients of the model like deformation and residual 
interactions on the GT strength distributions, which finally determine the 
decay half-lives, was also studied in the above references. In particular, the
sensitivity of the GT distributions to deformation has been used to learn about
the nuclear shapes when comparing with experiment \cite{expnacher}.

In this work, only allowed  $\beta^+/EC$ decays are considered. Forbidden transitions 
are in general much smaller and therefore, they can be safely neglected, especially 
in nuclei with small $Q_{EC}$-energies, such as those studied here. Allowed transitions 
correspond to $\Delta \pi =0$ and $\Delta J =0,\pm 1$ transitions and because 
of the small $Q_{EC}$ energies involved, only the low-lying excitations connecting 
proton with neutron states in the vicinity of the Fermi level obeying the above 
selection rules will contribute. In the case of the decay of even-even nuclei one 
has $0^+ \rightarrow 1^+$ transitions. In the case of odd-$A$ nuclei one needs the  
$J^{\pi}$ of the parent nucleus that will determine the allowed  $J^{\prime \pi}$ 
reached in the daughter nucleus. 

As mentioned earlier,
the spin and parity of the decaying nucleus are chosen by selecting the state 
occupied by the odd nucleon that minimizes the energy. However, other possibilities
of states in the neighborhood of the Fermi level different from the ground states
are also considered to study the $\beta^+/EC$-decay. This is because 
the calculations including deformation produce in many cases a high 
density of states around the Fermi level with a given ordering, which can be altered
by small changes in the theoretical treatment. In addition, fusion reactions
produce compound nuclei and subproducts in excited states that could decay directly.
Thus, it is interesting to know the $\beta^+/EC$-decay of the predicted 
ground states, as well as the decay of excited states close to it.
Since the $\beta^+/EC$-decay half-lives are sensitive to the spin-parity of the 
odd nucleon, calculations are performed for several states with opposite parity
close to the Fermi energy.

Comparison between $\alpha$- and $\beta^+/EC$-decay modes is crucial to understand
the possible branching and pathways of the original compound nucleus leading to 
stability. Unfortunately, not all the $\alpha$-decay half-lives of nuclei in this 
mass region have been measured yet. In those cases where experimental information 
on the total half-life and 
the percentage of the $\alpha$-decay mode intensity are available, $T_{\alpha}$ values 
are extracted and plotted in the following figures. To complete this information in 
other cases, the $\alpha$-decay half-lives have been estimated from phenomenological 
formulas that depend on the $Q_\alpha$ energies, using values taken from experiment 
\cite{audi_12} that can be seen in table \ref{table_q} with their uncertainties. 
Following the same approach as in \cite{sarri_19}, four different parametrizations 
are used, which have been fitted to account for the properties of SHN. Namely, they 
are the formula by Parkhomenko and Sobiczewski \cite{parkhomenko_05} (label 1 in the 
x-axis of the next figures), the  Royer formula \cite{royer_00} (label 2), and the 
Viola-Seaborg formula \cite{viola_66} with two different sets of parameters from 
\cite{parkhomenko_05} (label 3) and \cite{karpov_12,sobiczewski_89} (label 4). 

\section{Results and discussion}

\begin{figure*}[b]
\centering
\includegraphics[width=120mm]{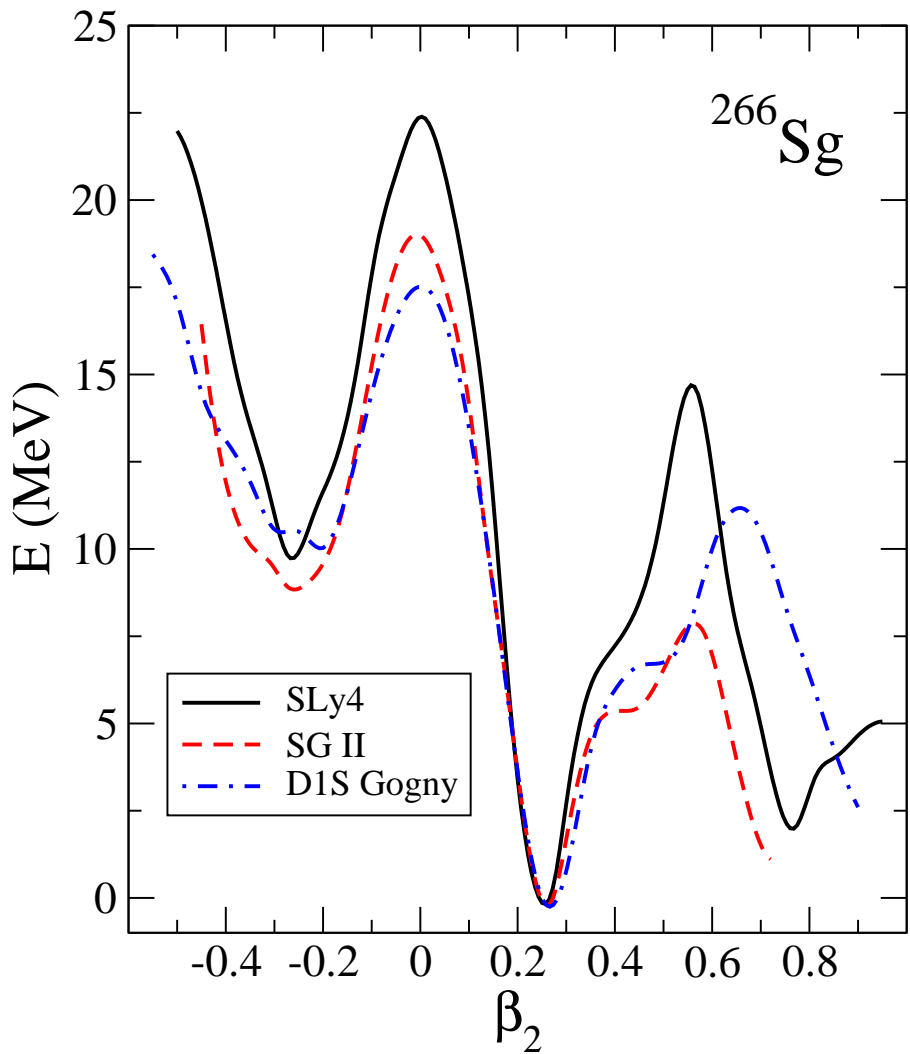}
\caption{Deformation-energy curves for the $^{266}$Sg isotope obtained from 
  constrained HF+BCS calculations with the Skyrme forces SLy4 \cite{chabanat}
  and SGII \cite{sg2}, as well as with the D1S Gogny \cite{gogny} interactions.}
\label{eq0}
\end{figure*}

DECs are first shown in figure \ref{eq0} for a selected 
isotope, $^{266}$Sg, which is representative of the nuclei in this mass region.
The energies in figure \ref{eq0} are relative to the ground state energy and are 
plotted as a function of the quadrupole deformation $\beta_2$. The results
corresponding to the interaction of reference in this work (SLy4) show a ground
state corresponding to a prolate shape at $\beta_2\approx 0.25$ and two 
excited configurations with oblate ($\beta_2\approx -0.25$) and superdeformed 
prolate ($\beta_2\approx 0.75$) shapes. The DECs obtained from other effective
interactions are quite similar and, in particular, the prolate deformation of
the ground state is very robust. To illustrate this feature, figure \ref{eq0}
shows also the results obtained with other standard Skyrme force, SGII
from \cite{sg2}, as well as the results obtained with the finite-range D1S
Gogny interaction \cite{gogny}.
The profile of the DEC turns out to be very similar to the DECs obtained for 
the other isotopes discussed in this work and agree also quite well with 
calculations performed with the Gogny D1S interaction \cite{gogny}.
In this work we calculate energy distributions of the GT strength and their 
corresponding half-lives for the ground state prolate configurations at 
$\beta_2\approx 0.25$.
Deformations of parent and daughter decay partners are assumed to be the
same in this work. This is a widely used approximation based on the spin-isospin
character of the Gamow-Teller operator that does not contain any radial dependency.
Thus, the spatial functions of parent and daughter wave functions are expected
to be as close as possible to overlap maximally. As a result, transitions
connecting different radial structures in the parent and daughter nuclei are
suppressed. This is justified by the fact that core polarization effects in the
daughter nuclei are negligible, as it can be seen, for example, from Gogny
calculations \cite{gogny}, where the DECs of parent and corresponding daughter
isotopes considered in this work are practically the same with ground states at
$\beta_2\approx 0.25$.
Consequently, given the small polarization effects and the suppression of the
overlaps with different deformations, only GT transitions between parent and
daughter partners with like deformations are considered in the present work.

In the next figures (figures \ref{fig_mdno} - \ref{fig_mt}) one can see the 
calculated half-lives $T_{\beta^+/EC}$ on the left-hand panels and $T_{\alpha}$ on the 
right-hand ones, grouped by isotopes. In the case of odd-$A$ nuclei, $T_{\beta^+/EC}$ 
are calculated for several $J^{\pi}$ values. For $T_{\alpha}$ there are calculations 
using the four models described above. The errors in the half-lives correspond to 
calculations using the extreme values of the $Q_{EC}$ and $Q_{\alpha}$ given by their 
experimental uncertainties in table \ref{table_q}.
Experimental data for $T_{\beta^+/EC}$ ($T_{\alpha}$) are extracted and shown under 
the label 'exp' in those cases where both the total experimental half-lives and 
the percentage of $\beta$ ($\alpha$) decay intensities are measured. 

Figure \ref{fig_mdno} contains the results for Md ($Z=101$) and No ($Z=102$) isotopes.
In this case the $\beta^+/EC$-decay half-lives are orders of magnitude larger than 
the corresponding $\alpha$-decay half-lives for a given isotope and have no
special interest.

Figure \ref{fig_lr} shows the results for the odd-$A$ Lr isotopes ($Z=103$). In this
case $J^{\pi}$ is determined by the odd-proton state. The experimental $J^{\pi}$ 
assignments \cite{audi_12} are mostly uncertain (values within parenthesis) 
or estimated from systematic trends in neighboring nuclei (denoted with \#). 
They are $J^{\pi}=(1/2^-)_{\rm g.s.}$ and $(7/2^-)$ in $^{255}$Lr,
$J^{\pi}=(1/2^-)$ in $^{257}$Lr, and $J^{\pi}=1/2^-$\# in $^{259}$Lr. 
In the present calculations, states $1/2^-$, $7/2^-$ and $7/2^+$ very close in energy 
around the Fermi surface are obtained, and then, calculations for these 
three possibilities are performed. 
The $\beta^+/EC$-decay half-lives of the positive-parity states turn out to be much
shorter than the corresponding half-lives of the negative-parity states.
The latter are always orders of magnitude larger than the $\alpha$-decay 
half-lives, but the former can compete in some instances with $\alpha$-decay.
Namely, in the case of the positive-parity states we obtain comparable $\beta^+/EC$-
and $\alpha$-decay half-lives for $^{255,263}$Lr isotopes. In the case of $^{261}$Lr 
the difference is within a factor of 10, whereas for $^{259}$Lr the difference
is already about two orders of magnitude.

Figure \ref{fig_rf} corresponds to Rf isotopes ($Z=104$). Even-even and 
odd-$N$ isotopes are considered. In the case of odd-$N$ nuclei, $J^{\pi}$ is determined 
by the odd neutron state. The systematics of the $J^{\pi}$ in the various isotopes is 
not clear since each isotope has a different number of neutrons and the last neutron 
occupies a different orbital. Experimentally, the $J^{\pi}$ assignments are as follows,
$J^{\pi}=(9/2^-)_{\rm g.s.}$ and $5/2^+$\# in $^{255}$Rf,
$J^{\pi}=(1/2^+)_{\rm g.s.}$ and $(11/2^-)$ in $^{257}$Rf,
$J^{\pi}=7/2^+\#_{\rm g.s.}$, $(3/2^+)$ and $(9/2^+)$ in $^{259}$Rf,
$J^{\pi}=3/2^+\#_{\rm g.s.}$ and $9/2^+$\# in $^{261}$Rf,
$J^{\pi}=3/2^+$\# in $^{263}$Rf, and $J^{\pi}=3/2^+$\# in $^{265}$Rf.
In the calculations, various possibilities for the odd-neutron states are obtained 
including those mentioned above. $\beta^+/EC$-decay half-lives 
are performed for representative $J^{\pi}$ of both positive and negative parities.
One can observe again that the $\beta^+/EC$-decay half-lives of the positive-parity
states are shorter than the corresponding half-lives of the negative-parity states.
For the positive-parity states, 
the half-lives of $\beta$- and $\alpha$-decays are similar in $^{255,265}$Rf.
In the case of $^{257}$Rf ($^{256}$Rf) the $T_{\beta^+/EC}$ are within a factor of 10 
(100) larger than $T_{\alpha}$.

In the case of the odd-$Z$ Db isotopes ($Z=105$) in figure \ref{fig_db}, the 
experimental $J^{\pi}$ is found to be $J^{\pi}=(9/2^+)_{\rm g.s.}$ and $(1/2^-)$ in 
$^{257}$Db and $J^{\pi}=(9/2^+)$ in $^{261}$Db. These states are found in the calculations
close to the Fermi surface, as well as $5/2^-$ states and half-lives are calculated 
for these options. Referring again to the positive-parity states that exhibit smaller 
half-lives than the negative ones, one can see that the half-lives of $\beta$- and 
$\alpha$-decays are comparable in $^{257,265,267}$Db, while there is one order of 
magnitude of difference in $^{263}$Db and two orders in $^{259,261}$Db.

Next we consider Sg ($Z=106$) isotopes in figure \ref{fig_sg}. The spin-parity 
experimental assignments of the odd isotopes are as follows:
$J^{\pi}=1/2^+$\# in $^{259}$Sg, 
$J^{\pi}=(3/2^+)_{\rm g.s.}$ and $(11/2^-)$ in $^{261}$Sg,
$J^{\pi}=7/2^+\#_{\rm g.s.}$ and $3/2^+$\# in $^{263}$Sg, and
$J^{\pi}=9/2^+\#_{\rm g.s.}$ and $3/2^+$\# in $^{265}$Sg.
In this case the half-lives $T_{\beta^+/EC}$ are quite similar for both parities
in the lighter isotopes and they start to diverge from $^{267}$Sg.
The $\beta$- and $\alpha$-decay half-lives are comparable in $^{267}$Sg, whereas they
differ by about one order of magnitude in $^{265}$Sg and by two orders in $^{258,263,266}$Sg.

The half-lives of Bh isotopes ($Z=107$) are shown in figure \ref{fig_bh}. 
The only experimental value of $J^{\pi}$ assigned is $(5/2^-)$ in $^{261}$Bh.
The calculations give $5/2^-$ and $9/2^+$ states at the Fermi level and $\beta$-decay 
calculations are made for both of them. In the lighter isotopes considered there is
not big difference between the $\beta^+/EC$-decay half-lives calculated with the 
different spin-parities. For isotopes heavier than $N=162$ the difference is much
larger. The $\beta$- and $\alpha$-decay half-lives are comparable in 
$^{269}$Bh, they differ by one order of magnitude in $^{267}$Bh and by two orders of
magnitude in $^{265}$Bh.

Even-even isotopes of Hs ($Z=108$) are shown in figure \ref{fig_hs}. The spin-parity
experimental assignments are $7/2^+$\# for $^{263}$Hs, $3/2^+\#_{\rm g.s.}$ and $9/2^+$\#
for $^{265}$Hs, $5/2^+$\# for $^{267}$Hs, $9/2^+$\# for $^{269}$Hs, and $3/2^+$\# for 
$^{273}$Hs. Similar to the case of Sg isotopes, the $\beta^+/EC$-decay half-lives 
calculated with positive-parity states are very close to those from negative-parity
states in the lighter isotopes up to $N=159$. Heavier isotopes show a more
drastic dependence on the parity of the states. The $\beta$- and $\alpha$-decay
half-lives 
differ by less than a factor of 10 in $^{269}$Hs and about two orders of magnitude in 
$^{267,268}$Hs.

Finally, in Mt isotopes ($Z=109$), the half-lives plotted in figure \ref{fig_mt}
show that the difference between $T_{\alpha}$ and  $T_{\beta^+/EC}$ is more than two
orders of magnitude in all isotopes except $^{271}$Mt, where it is about two orders. 

From the above figures one can also learn about the uncertainties associated with
different aspects of the calculations, in particular with the $Q_{EC}$ energies 
that are plotted as error bars and with the $J^{\pi}$ assignments in the case of
odd-$A$ isotopes.

Tables \ref{table2}-\ref{table4} summarize the main results obtained in this work
regarding the comparison between the half-lives of the $\beta^+/EC$- and $\alpha$-decay
modes. Table \ref{table2} contains the calculated half-lives for the isotopes with
comparable $T_{\beta^+/EC}$ and $T_{\alpha}$. Experimental
values extracted from \cite{audi_12} are also shown when available. In the case 
of $T_{\alpha}$ the values shown correspond to the average value obtained from the
four formulas considered in this work. SF in this mass region is
another possible decay mode that might also compete with $\alpha$ and $\beta$ decay
in some isotopes. For comparison, the available experimental half-lives for SF,
$T_{SF}$, obtained from the total half-lives and percentage of the SF
decay mode intensity from \cite{audi_12} are quoted in the last column of
the table. 

Table \ref{table3} shows the same information as in table \ref{table2}, but for
the isotopes whose $T_{\beta^+/EC}$ 
and $T_{\alpha}$ values are within a factor of ten. Similarly, table \ref{table4} 
contains the information on the isotopes with $T_{\beta^+/EC}$ and $T_{\alpha}$
differing by about two orders of magnitude.

\begin{figure*}[htb]
\centering
\includegraphics[width=150mm]{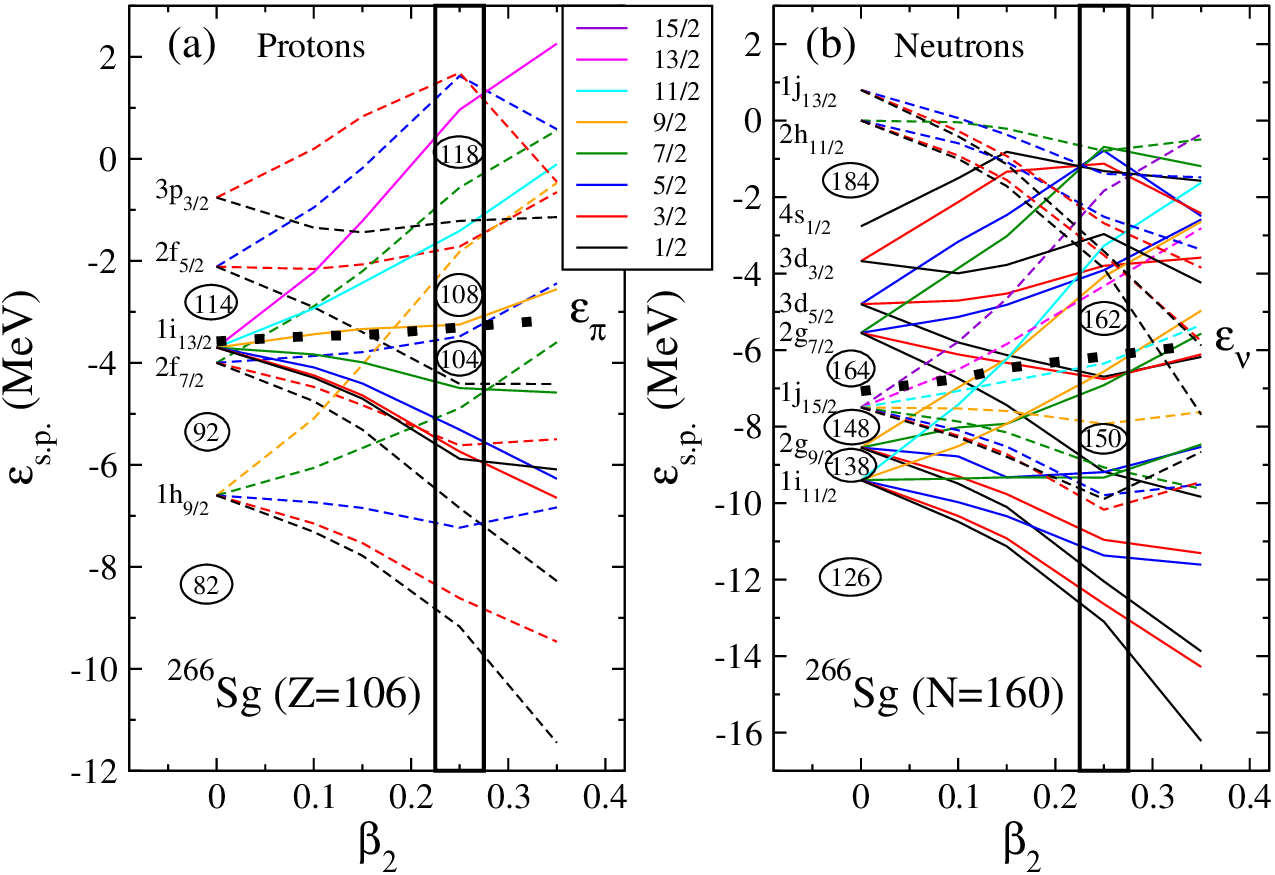}
\caption{ Single-particle energies obtained with SLy4 for (a) protons and (b) 
neutrons in $^{266}$Sg as a function of the quadrupole deformation $\beta_2$. 
The Fermi levels for protons ($\varepsilon_\pi$) and neutrons ($\varepsilon_\nu$) are
depicted as thick dotted black lines. Positive-parity states are shown with solid
lines, whereas negative-parity states are shown with dashed lines. }
\label{fig_nilsson}
\end{figure*}

To understand why in the case of odd-$A$ isotopes the $T_{\beta^+/EC}$ for 
positive-parity states are always significantly lower than the same half-lives for 
negative-parity states, one has to analyze the different scenarios for the decay
when the odd nucleon in the parent nucleus has a positive or a negative parity. 
This parity will determine the parity of all the states reached in the daughter 
nucleus (allowed transitions). For that purpose figure \ref{fig_nilsson} shows a 
Nilsson-like diagram, where the single-particle energies are plotted as a function 
of the quadrupole deformation $\beta_2$ for protons (left) and neutrons (right) 
in the case of $^{266}$Sg with $Z=106$ and $N=160$. 
The calculations correspond to the Skyrme interaction SLy4. Fermi levels for protons  
($\varepsilon_\pi$) and neutrons ($\varepsilon_\nu$) are plotted as thick dotted black 
lines. Positive-parity states are shown with solid lines, whereas negative-parity states
are shown with dashed lines. The spherical shells are shown at $\beta_2=0$ with their 
spherical quantum numbers. The boxes centered at $\beta_2=0.25$ correspond to the 
regions of interest determined by the quadrupole deformation of the equilibrium 
ground state in this mass region. The color code used 
to plot the different components of the angular momenta is also shown.

The spherical shells involved in the $\beta^+/EC$-decay in increasing order of energy 
are the following: 
$h_{9/2},\ f_{7/2},\ i_{13/2},\ f_{5/2},\ p_{3/2}$ for protons and 
$i_{11/2},\ g_{9/2},\ j_{15/2},\ g_{7/2},\ d_{5/2},\ d_{3/2},\ s_{1/2},\ h_{11/2},\ j_{13/2}$ 
for neutrons.
In this analysis it is sufficient to focus on the states in the vicinity of the Fermi 
energy because only low-lying transitions involving the odd nucleon are relevant to
calculate half-lives in nuclei with low $Q$-energies, which is the case of the isotopes 
studied here.
In the $\beta^+/EC$-decay one proton is transformed into one neutron. In the case
of odd-proton isotopes, the odd-proton is directly involved in the low-lying 
transitions below the $Q$-window that determine the $\beta^+/EC$-decay half-lives.
In the case of odd-neutron isotopes, the low-lying GT excitations corresponding to
$\beta^+/EC$-decay involve proton states in the vicinity of the Fermi level 
that match the allowed quantum-numbers given by the odd neutron.

Focusing on the proton single-particle energies within the box around $\beta_2=0.25$, 
one can see that among the odd-proton isotopes considered, Lr ($Z=103$) would have 
the odd proton placed in one of the orbitals $7/2^-$, $7/2^+$, and $1/2^-$ close to the 
Fermi energy. They originate in the spherical shells $h_{9/2},\ i_{13/2},\ f_{5/2}$, 
respectively. In the case of Db ($Z=105$) and Bh ($Z=107$), the states involved 
would be $5/2^-\, (f_{7/2})$ and $9/2^+\, (i_{13/2})$, whereas in the case of Mt($Z=109$) 
one finds the states  $9/2^+\, (i_{13/2})$, $9/2^-\, (h_{9/2})$, $3/2^-\, (f_{5/2})$, and 
$11/2^+\, (i_{13/2})$ close to the Fermi level. All of above states have been considered 
in the decay of these odd-$A$ isotopes. Therefore, in the odd-proton isotopes, the 
states mentioned above would decay into neutron states in the vicinity of the neutron
Fermi energy, which is shown in the right panel of figure \ref{fig_nilsson} within the 
black box centered at $\beta_2=0.25$. For neutron numbers between $N$=150-162 the states 
involved are $9/2^-\, (j_{15/2})$,  $7/2^+\, (g_{9/2})$, $9/2^+\, (i_{11/2})$, 
$1/2^+\, (d_{5/2})$, $3/2^+\, (g_{7/2})$, and  $11/2^-\, (j_{15/2})$. 
Beyond $N$=162, new states appear with  $13/2^-\, (j_{15/2})$, $9/2^+\, (g_{9/2})$, 
$5/2^+\, (g_{7/2})$,  $3/2^+\, (d_{5/2})$, $3/2^-\, (h_{11/2})$, and  
$1/2^-\, (h_{11/2},j_{13/2})$.

Then, it is easy to understand that the transitions involving the positive-parity 
proton states, namely  $7/2^+$ and $9/2^+$, would match the neutron states 
$5/2^+$, $7/2^+$, and $9/2^+$, while the negative-parity proton states $5/2^-$ cannot 
match the $9/2^-$, $11/2^-$ or $13/2^-$ and only in the heavier isotopes the $9/2^-$ 
proton states states can match the $9/2^-$ and $11/2^-$ neutron states.
This explains qualitatively why the decays from even-parity states are much faster
than the decays from odd-parity states.
In the case of odd-neutron isotopes, the argument is similar, but now, the odd 
neutron in the parent nucleus determines the proton states involved in the 
transitions. 

The comparison of the calculations with the available experimental half-lives 
in both cases $T_{\beta^+/EC}$ and $T_{\alpha}$ is in general quite satisfactory,
which helps to be confident in the reliability of the calculations.
More specifically, in the case of $\beta^+/EC$-decay, the experimental $T_{\beta^+/EC}$
in $^{255}$Lr lies within the calculated half-lives with positive- or negative-parity
states, whereas in the case of $^{255}$Lr and $^{257}$Rf the experiment is very close 
to the calculation with the $7/2^+$ state. 
On the other hand, in $^{257}$Db, the experiment is closer to the calculation with
negative parity and in $^{263}$Db the experiment lies between the predictions with
positive or negative parities. Finally, in $^{261}$Sg the experiment is close to
the calculations with both parities.
In the case of $\alpha$-decay the experimental information on half-lives is more
abundant. In general, the predictions of the different formulas considered agree 
within one order of magnitude with the measurements.

\section{Conclusions}

In this work, the $\beta^+/EC$-decay half-lives of some selected even-even and 
odd-$A$ isotopes in the region $Z=101-109$ and $N=151-168$ are studied. 
Namely, $^{259}$Md, $^{260,261}$No, odd $^{255-265}$Lr, $^{255-267}$Rf, odd $^{257-269}$Db,
$^{258-271}$Sg, odd $^{261-273}$Bh, $^{263-275}$Hs, and odd $^{265-277}$Mt. 
The microscopic formalism used to describe the nuclear structure of the decay
partners is based on a deformed Skyrme HF+BCS approach.

Uncertainties in the experimental $Q_{EC}$ energies are translated into uncertainties
of the half-lives calculated with them. In the case of odd-$A$ nuclei, different 
$J^{\pi}$ assignments are considered to learn about their influence on the final
half-lives. It is found that in odd-$A$ isotopes the $T_{\beta^+/EC}$ for the
positive-parity states are always shorter than the  $T_{\beta^+/EC}$ for the
negative-parity states.
The results for $T_{\beta^+/EC}$ are compared with the $\alpha$-decay 
half-lives $T_{\alpha}$ obtained from phenomenological formulas using experimental
$Q_{\alpha}$ energies and their uncertainties. The agreement between the calculated 
and the available experimental half-lives in both cases $T_{\beta^+/EC}$ and 
$T_{\alpha}$ is found to be always within a factor of 10, granting the category 
of trustable predictions.

$T_{\alpha}$ are in most cases lower than the corresponding $T_{\beta^+/EC}$ for a
given isotope. This difference is about two orders of magnitude in  $^{259}$Lr,
$^{256}$Rf, $^{259,261}$Db, $^{258,263,266}$Sg, $^{265}$Bh, $^{267,268}$Hs, and $^{271}$Mt.
In the cases of $^{261}$Lr, $^{257}$Rf, $^{263}$Db, $^{265}$Sg and $^{267}$Bh the 
difference is only about one order of magnitude. Finally, the isotopes $^{255,263}$Lr, 
$^{255,265}$Rf, $^{257,265,267}$Db,  $^{267}$Sg, $^{269}$Bh, and $^{269}$Hs have comparable 
values of the half-lives for the $\beta^+/EC$- and $\alpha$-decay modes. Therefore, 
these different modes will compete in the latter cases favoring new branches of 
decay in the $\beta^+/EC$ direction that have not yet been sufficiently studied. 
This opens new possibilities
to reach unexplored roads towards the predicted islands of stability.

\begin{acknowledgments}
%\ack
%\section{Acknowledgments}
I would like to thank G. G. Adamian for the suggestion of this problem and 
for useful discussions and valuable advice. 
This work was supported by Ministerio de Ciencia e Innovaci\'on 
MCI/AEI/FEDER,UE (Spain) under Contract No. PGC2018-093636-B-I00.  
%\end{acknowledgments}
\end{acknowledgments}

%\clearpage
%\section*{References}

%%%%%%%%%%%%%%%%%%%%%%%%%%%%%%%%%%%%%%%%%%%%%%%%%%%%%%%%%%%%%%%%%%%%%%%%%%%%%%%%%%%%%%

%\clearpage

\begin{table*}[htb]
%\small
\caption{Experimental $Q_{\rm EC}$ and $Q_\alpha$
energies (MeV) from AME2012 \cite{audi_12}}
\scalebox{0.7}{
{\begin{tabular}{ccc|ccc|ccc} \hline \hline
Nucleus  &  $Q_{\rm EC}$ & $Q_\alpha$ & Nucleus  &  $Q_{\rm EC}$ & $Q_\alpha$ & Nucleus  &  $Q_{\rm EC}$ & $Q_\alpha$ \\
\hline
$^{259}$Md   &  0.0  $\pm$ 0.3  &  7.11  $\pm$ 0.20  &&&&&& \\
&&&&&&&& \\
$^{260}$No   & -0.9  $\pm$ 0.4  &  7.70  $\pm$ 0.20  &
$^{261}$No   &  0.0  $\pm$ 0.5  &  7.44  $\pm$ 0.20  &&& \\
&&&&&&&& \\
$^{255}$Lr   & 3.140 $\pm$ 0.023&  8.556 $\pm$ 0.007 &
$^{257}$Lr   &  2.36 $\pm$ 0.05 &  9.01  $\pm$ 0.03  &
$^{259}$Lr   &  1.74 $\pm$ 0.12 &  8.58  $\pm$ 0.07    \\
$^{261}$Lr   &  1.11 $\pm$ 0.28 &  8.14  $\pm$ 0.20  &
$^{263}$Lr   &  0.60 $\pm$ 0.57 &  7.68  $\pm$ 0.20  &
$^{265}$Lr   &         -        &  7.23  $\pm$ 0.20    \\
&&&&&&&& \\
$^{255}$Rf   & 4.38  $\pm$ 0.12 &  9.055 $\pm$ 0.004 &
$^{256}$Rf   & 2.48  $\pm$ 0.08 &  8.926 $\pm$ 0.015 &
$^{257}$Rf   & 3.26  $\pm$ 0.05 &  9.083 $\pm$ 0.008   \\
$^{258}$Rf   & 1.56  $\pm$ 0.11 &  9.19  $\pm$ 0.03  &
$^{259}$Rf   & 2.51  $\pm$ 0.10 &  9.13  $\pm$ 0.07  &
$^{260}$Rf   & 0.87  $\pm$ 0.24 &  8.90  $\pm$ 0.20    \\
$^{261}$Rf   & 1.76  $\pm$ 0.21 &  8.65  $\pm$ 0.05  &
$^{262}$Rf   & 0.29  $\pm$ 0.30 &  8.49  $\pm$ 0.20  &
$^{263}$Rf   & 1.06  $\pm$ 0.34 &  8.25  $\pm$ 0.15    \\
$^{264}$Rf   & -0.20 $\pm$ 0.60 &  8.04  $\pm$ 0.30  &
$^{265}$Rf   &  0.46 $\pm$ 0.71 &  7.81  $\pm$ 0.30  &
$^{266}$Rf   & -1.5  $\pm$ 0.7  &  7.55  $\pm$ 0.30  \\
$^{267}$Rf   &       -          &  7.89  $\pm$ 0.30  & &&&&& \\
&&&&&&&& \\
$^{257}$Db   &  4.34 $\pm$ 0.20 &  9.026 $\pm$ 0.020 &
$^{259}$Db   &  3.63 $\pm$ 0.09 &  9.62  $\pm$ 0.05  &
$^{261}$Db   &  2.93 $\pm$ 0.12 &  9.22  $\pm$ 0.10    \\
$^{263}$Db   &  2.32 $\pm$ 0.25 &  8.83  $\pm$ 0.15  &
$^{265}$Db   &  1.80 $\pm$ 0.42 &  8.50  $\pm$ 0.10  &
$^{267}$Db   &  0.63 $\pm$ 0.71 &  7.92  $\pm$ 0.30    \\
$^{269}$Db   &          -       &  8.49  $\pm$ 0.30  & &&&&& \\
&&&&&&&& \\
$^{258}$Sg   &  3.45 $\pm$ 0.51 &  9.62  $\pm$ 0.30  &
$^{259}$Sg   &  4.57 $\pm$ 0.13 &  9.804 $\pm$ 0.021 &
$^{260}$Sg   &  2.88 $\pm$ 0.10 &  9.901 $\pm$ 0.010   \\
$^{261}$Sg   &  3.76 $\pm$ 0.11 &  9.714 $\pm$ 0.015 &
$^{262}$Sg   &  2.11 $\pm$ 0.15 &  9.600 $\pm$ 0.015 &
$^{263}$Sg   &  3.08 $\pm$ 0.19 &  9.40  $\pm$ 0.06    \\
$^{264}$Sg   &  1.42 $\pm$ 0.37 &  9.21  $\pm$ 0.20  &
$^{265}$Sg   &  2.31 $\pm$ 0.26 &  9.05  $\pm$ 0.11  &
$^{266}$Sg   &  0.88 $\pm$ 0.37 &  8.80  $\pm$ 0.10    \\
$^{267}$Sg   &  1.76 $\pm$ 0.50 &  8.63  $\pm$ 0.21  &
$^{268}$Sg   & -0.2  $\pm$ 0.7  &  8.30  $\pm$ 0.30  &
$^{269}$Sg   &  0.67 $\pm$ 0.77 &  8.70  $\pm$ 0.05    \\
$^{270}$Sg   & -0.7  $\pm$ 0.8  &  8.99  $\pm$ 0.30  &
$^{271}$Sg   &      -           &  8.89  $\pm$ 0.11    \\
&&&&&&&& \\
$^{261}$Bh   &  5.13 $\pm$ 0.21 &  10.50 $\pm$ 0.05  &
$^{263}$Bh   &  4.31 $\pm$ 0.32 &  10.08 $\pm$ 0.30  &
$^{265}$Bh   &  3.56 $\pm$ 0.26 &  9.68  $\pm$ 0.21    \\
$^{267}$Bh   &  2.93 $\pm$ 0.38 &  9.23  $\pm$ 0.20  &
$^{269}$Bh   &  1.67 $\pm$ 0.52 &  8.57  $\pm$ 0.30  &
$^{271}$Bh   &  1.23 $\pm$ 0.73 &  9.49  $\pm$ 0.16    \\
$^{273}$Bh   &  0.62 $\pm$ 0.90 &  9.06  $\pm$ 0.30  &&&&&& \\
&&&&&&&& \\
$^{263}$Hs   &  5.22 $\pm$ 0.33 & 10.73  $\pm$ 0.05  &
$^{264}$Hs   &  3.51 $\pm$ 0.18 & 10.59  $\pm$ 0.02  &
$^{265}$Hs   &  4.55 $\pm$ 0.24 & 10.47  $\pm$ 0.11    \\
$^{266}$Hs   &  3.03 $\pm$ 0.17 & 10.346 $\pm$ 0.016 &
$^{267}$Hs   &  3.89 $\pm$ 0.28 & 10.037 $\pm$ 0.013 &
$^{268}$Hs   &  2.02 $\pm$ 0.48 &  9.623 $\pm$ 0.016   \\
$^{269}$Hs   &  3.11 $\pm$ 0.39 &  9.37  $\pm$ 0.16  &
$^{270}$Hs   &  0.86 $\pm$ 0.38 &  9.05  $\pm$ 0.04  &
$^{271}$Hs   &  1.78 $\pm$ 0.53 &  9.51  $\pm$ 0.11    \\
$^{272}$Hs   &  0.22 $\pm$ 0.74 &  9.78  $\pm$ 0.20  &
$^{273}$Hs   &  1.34 $\pm$ 0.83 &  9.73  $\pm$ 0.05  &
$^{274}$Hs   & -0.1  $\pm$ 0.9  &  9.57  $\pm$ 0.20    \\
$^{275}$Hs   &  0.93 $\pm$ 0.84 & 9.44   $\pm$ 0.05  &&&&&& \\
&&&&&&&& \\
$^{265}$Mt   &  5.78 $\pm$ 0.45 &  11.12 $\pm$ 0.40  &
$^{267}$Mt   &  5.14 $\pm$ 0.51 &  10.87 $\pm$ 0.40  &
$^{269}$Mt   &  4.72 $\pm$ 0.48 &  10.53 $\pm$ 0.40    \\
$^{271}$Mt   &  3.33 $\pm$ 0.44 &  9.91  $\pm$ 0.20  &
$^{273}$Mt   &  2.53 $\pm$ 0.60 & 10.60  $\pm$ 0.30  &
$^{275}$Mt   &  2.01 $\pm$ 0.75 &  10.21 $\pm$ 0.15    \\
$^{277}$Mt   &  1.28 $\pm$ 0.94 &  9.71  $\pm$ 0.20  &&&&&& \\
\hline \hline
\label{table_q}
\end{tabular}}
}
\end{table*}

%%%%%%%%%%%%%%%%%%%%%%%%%%%%%%%%%%%%%%%%%%%%%%%%%%%%%%%%%%%%%%%%%%%%%%%%%%%%%%%%%%%%%%

\begin{table*}[htb]
\caption{Experimental \cite{audi_12} and calculated  $T_{\beta^+/EC}$ [s] and 
  $T_{\alpha}$ [s] for isotopes with comparable half-lives. Experimental half-lives
  for spontaneous fission  \cite{audi_12},  $T_{SF}$ [s], are also shown for
comparison.}
%\begin{indented}
%\item[]
{\begin{tabular}{cccccccc} \hline
    Nucleus  &  \multicolumn{2}{c}{$T_{\beta^+/EC}$} && \multicolumn{2}{c} {$T_{\alpha}$}
    &&  $T_{SF}$ \\
\cline{2-3} \cline{5-6}
& exp. & calc. && exp. & calc. && exp.\\ 
\hline
\\
$^{255}$Lr   &  120 & 42.6   && 42  & 28.6  &&        \\
$^{263}$Lr   &      & 50231  &&     & 30925 &&        \\
$^{255}$Rf   &      &  7.2   && 3.5 & 1.9   && 3.2    \\
$^{265}$Rf   &      & 28211  &&     & 25260 && 396    \\
$^{257}$Db   &  230 & 4.1    && 2.5 & 5.4   && 46     \\
$^{265}$Db   &      & 165    &&     & 233   &&        \\
$^{267}$Db   &      & 6192   &&     & 24700 && 16560  \\
$^{267}$Sg   &      & 254    && 636 & 206   &&  130   \\
$^{269}$Bh   &      & 192    &&     & 770   &&        \\
$^{269}$Hs   &      & 26.1  && 27   & 5.6   &&        \\
\hline \hline 
\label{table2}
\end{tabular}}
%\end{indented}
\end{table*}

%%%%%%%%%%%%%%%%%%%%%%%%%%%%%%%%%%%%%%%%%%%%%%%%%%%%%%%%%%%%%%%%%%%%%%%%%%%%%%%%%%%%%%

\begin{table*}[htb]
\caption{Same as in Table \ref{table2}, but for isotopes with  $T_{\beta^+/EC}$ and 
$T_{\alpha}$ half-lives differing by about one order of magnitude.}
%\begin{indented}
%\item[]
{\begin{tabular}{cccccccc} \hline 
Nucleus  &  \multicolumn{2}{c}{$T_{\beta^+/EC}$} && \multicolumn{2}{c} {$T_{\alpha}$}
   &&  $T_{SF}$ \\
\cline{2-3} \cline{5-6}
& exp. & calc. && exp. & calc. && exp.\\ 
\hline
\\
$^{261}$Lr   &      & 5414  &&      & 670  &&      \\
$^{257}$Rf   & 25.4 & 14.3  && 6.0  & 1.6  && 371  \\
$^{263}$Db   & 420  & 127   && 78.4 & 20.4 && 51.8 \\
$^{265}$Sg   &      & 128   && 18.4 & 9.9  && 184  \\
$^{267}$Bh   &      & 44.8  &&  22  & 6.4  &&      \\
\hline \hline
\label{table3}
\end{tabular}}
%\end{indented}
\end{table*}

%%%%%%%%%%%%%%%%%%%%%%%%%%%%%%%%%%%%%%%%%%%%%%%%%%%%%%%%%%%%%%%%%%%%%%%%%%%%%%%%%%%%%%

\begin{table*}[htb]
\caption{Same as in Table \ref{table2}, but for isotopes with  $T_{\beta^+/EC}$ and 
$T_{\alpha}$ half-lives differing by about two orders of magnitude.}
%\begin{indented}
%\item[]
{\begin{tabular}{cccccccc} \hline
Nucleus  &  \multicolumn{2}{c}{$T_{\beta^+/EC}$} && \multicolumn{2}{c} {$T_{\alpha}$} 
&&  $T_{SF}$ \\
\cline{2-3} \cline{5-6}
& exp. & calc. && exp. & calc. && exp. \\ 
\hline
\\
$^{259}$Lr   & 1033 & 2609  && 8.0  & 23.1  && 28.2  \\
$^{256}$Rf   &      & 79.1  &&      & 1.6   && 0.007 \\
$^{259}$Db   &      & 10.7  && 0.5  & 0.10  &&       \\
$^{261}$Db   &      & 40.1  && 16.7 & 1.4   && 6.2   \\
$^{258}$Sg   &      & 15.6  && 0.14 & 0.08  && 0.034 \\
$^{263}$Sg   &      & 58.8  && 1.1  & 0.9   &&  7.2  \\
$^{266}$Sg   &      & 1483  &&      & 19.1  &&  0.46 \\
$^{265}$Bh   &      & 16.9  && 1.2  & 0.33  &&       \\
$^{267}$Hs   &      & 15.5  && 0.07 & 0.08  &&  0.28 \\
$^{268}$Hs   &      & 86.9  && 1.4  & 0.36  &&       \\
$^{271}$Mt   &      & 28.6  && 0.4  & 0.36  &&       \\
\hline \hline
\label{table4}
\end{tabular}}
%\end{indented}
\end{table*}

%\clearpage

%%%%%%%%%%%%%%%%%%%%%%%%%%%%Fig1%%%%%%%%%%%%%%%%%%%%%%%%%%%%%%%%%%%%%%%%%%%%%%%%%%%%%%%%

%%%%%%%%%%%%%%%%%%%%%%%%%%%%Fig2%%%%%%%%%%%%%%%%%%%%%%%%%%%%%%%%%%%%%%%%%%%%%%%%%%%%%%%%
\begin{figure*}[htb]
\centering
\includegraphics[width=150mm]{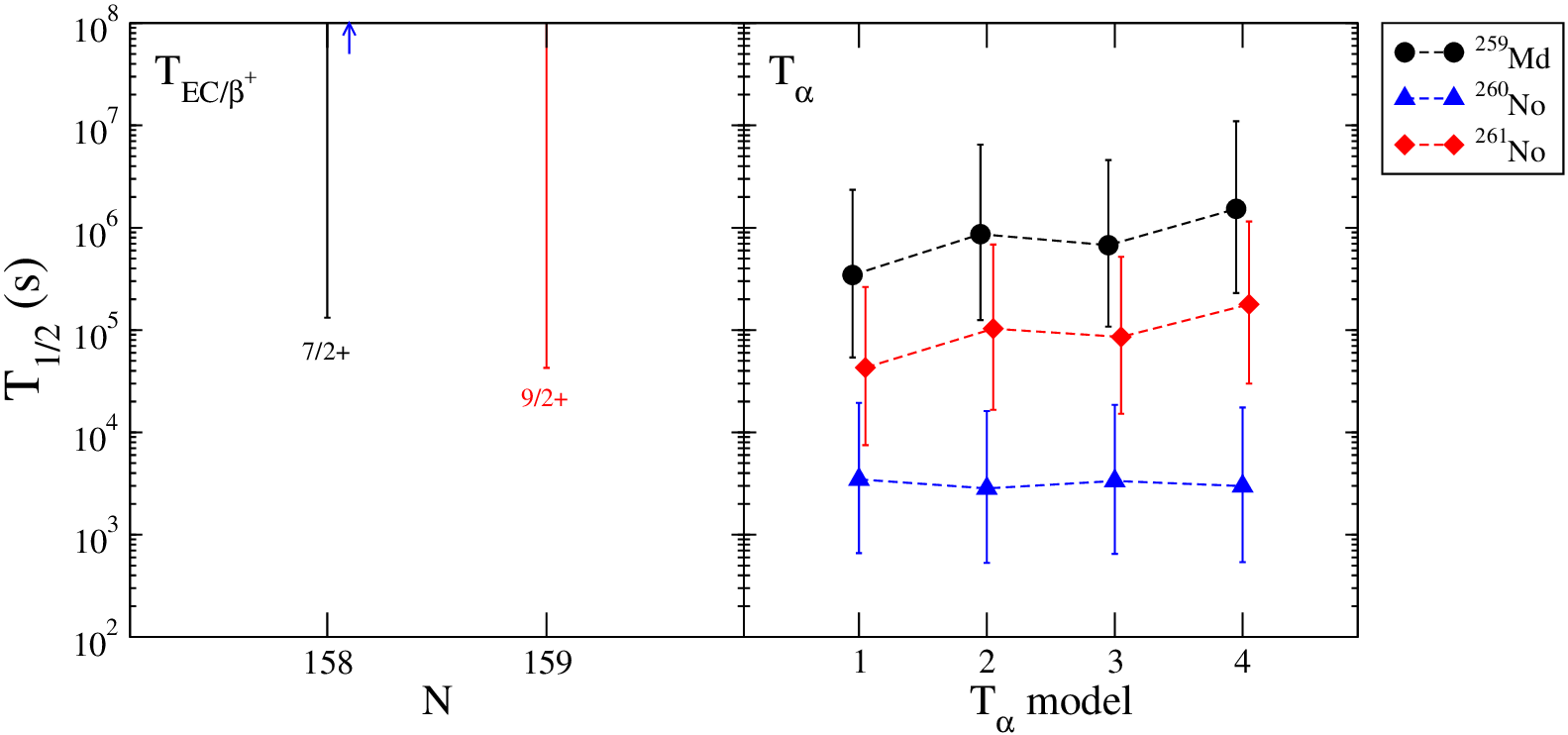}
\caption{Calculated half-lives for $^{259}$Md,  $^{260}$No, and $^{261}$No isotopes.
(left) Microscopic calculations of $T_{\beta^+/EC}$ for the ground state prolate
configurations. $J^\pi$ is indicated for odd-$A$ isotopes. 
(right) $T_{\alpha}$ from four different phenomenological formulas, labeled from 1 
up to 4 in the x axis (see text). $T_{1/2}$ in the y-axis is a short notation for
both $T_{\beta^+/EC}$ and  $T_{\alpha}$.}
\label{fig_mdno}
\end{figure*}

%%%%%%%%%%%%%%%%%%%%%%%%%%%%Fig3%%%%%%%%%%%%%%%%%%%%%%%%%%%%%%%%%%%%%%%%%%%%%%%%%%%%%%%%
\begin{figure*}[htb]
\centering
\includegraphics[width=150mm]{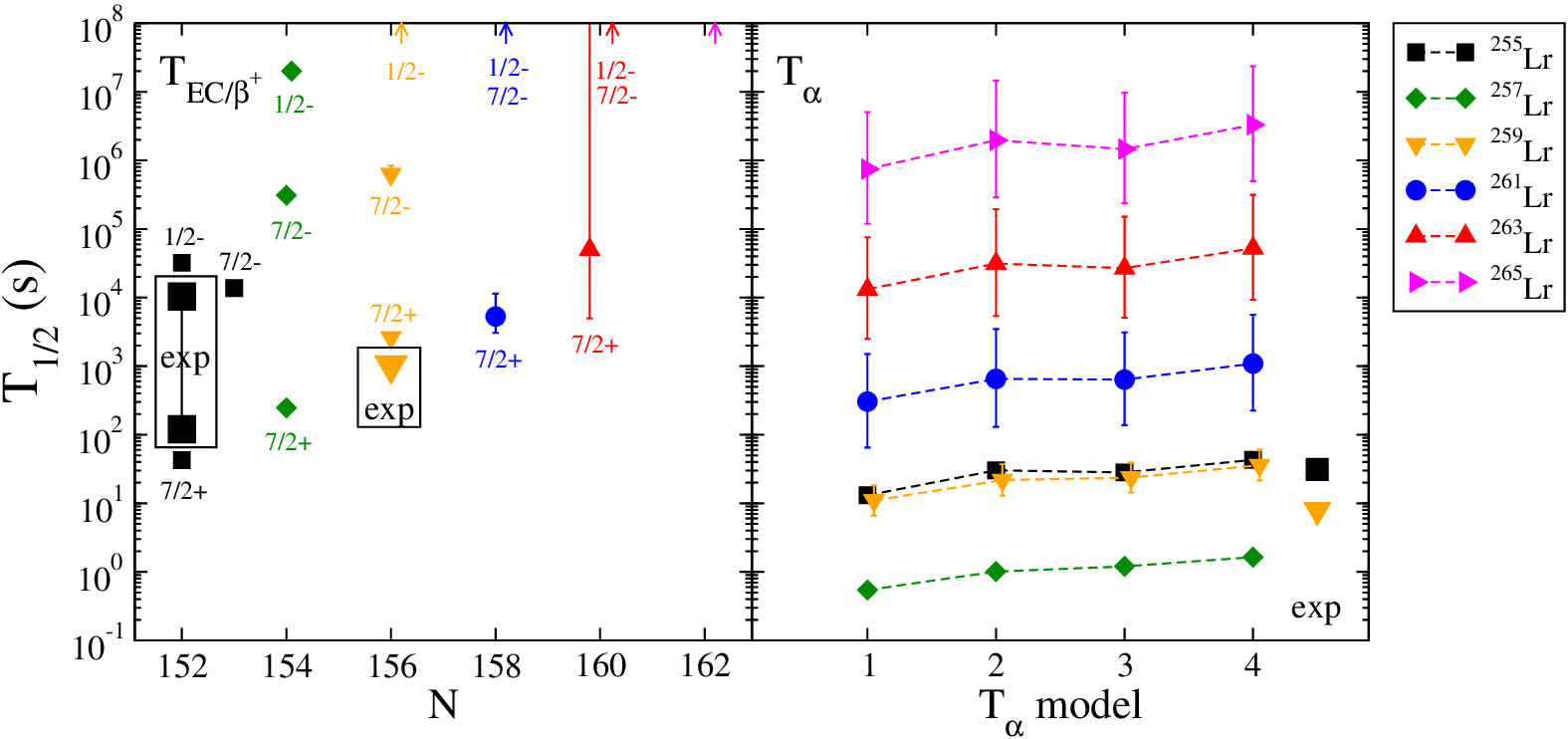}
\caption{Same as in figure \ref{fig_mdno}, but for the odd-$A$ Lr isotopes 
from $^{255}$Lr up to  $^{265}$Lr.}
\label{fig_lr}
\end{figure*}

%%%%%%%%%%%%%%%%%%%%%%%%%%%%Fig4%%%%%%%%%%%%%%%%%%%%%%%%%%%%%%%%%%%%%%%%%%%%%%%%%%%%%%%%
\begin{figure*}[htb]
\centering
\includegraphics[width=150mm]{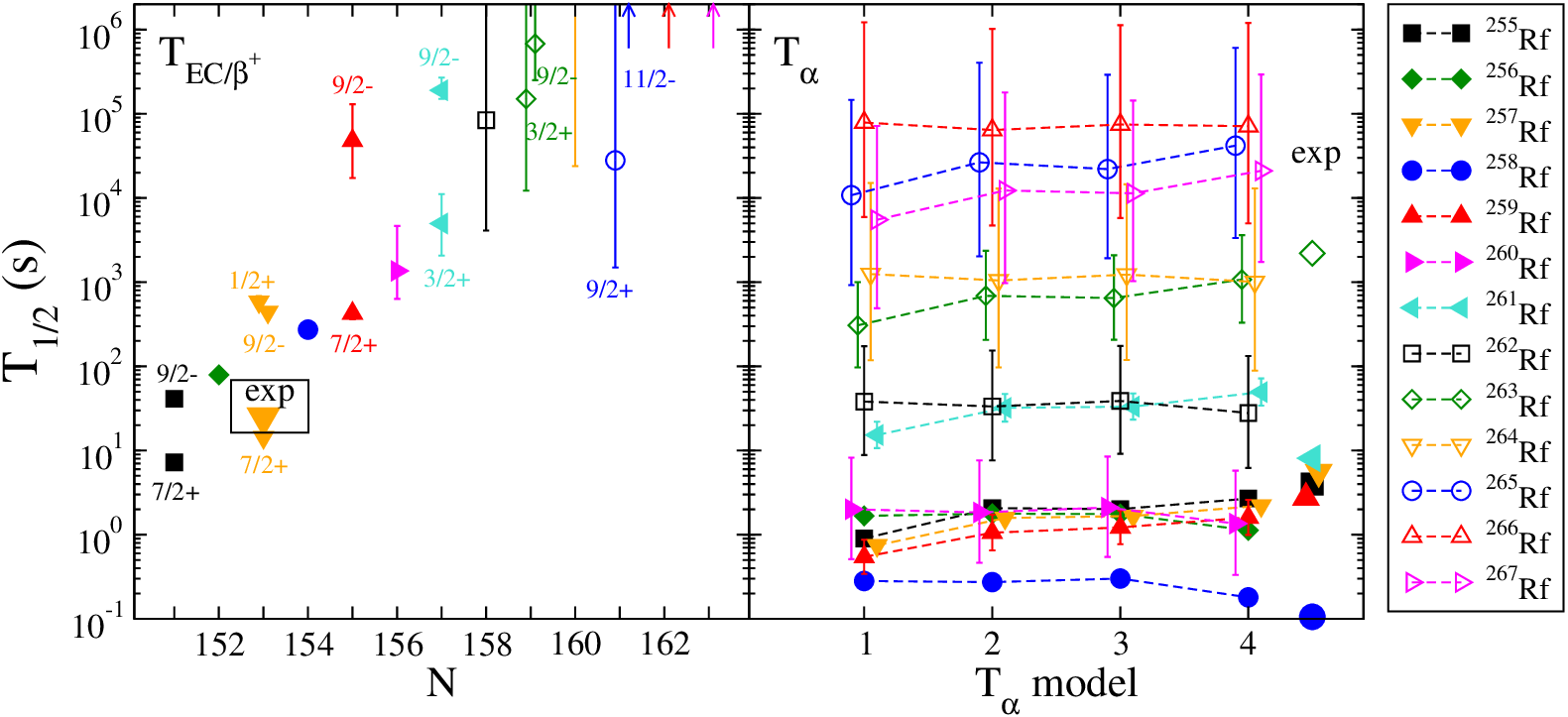}
\caption{Same as in figure \ref{fig_mdno}, but for $^{255-267}$Rf isotopes.}
\label{fig_rf}
\end{figure*}

%%%%%%%%%%%%%%%%%%%%%%%%%%%%Fig5%%%%%%%%%%%%%%%%%%%%%%%%%%%%%%%%%%%%%%%%%%%%%%%%%%%%%%%%
\begin{figure*}[htb]
\centering
\includegraphics[width=150mm]{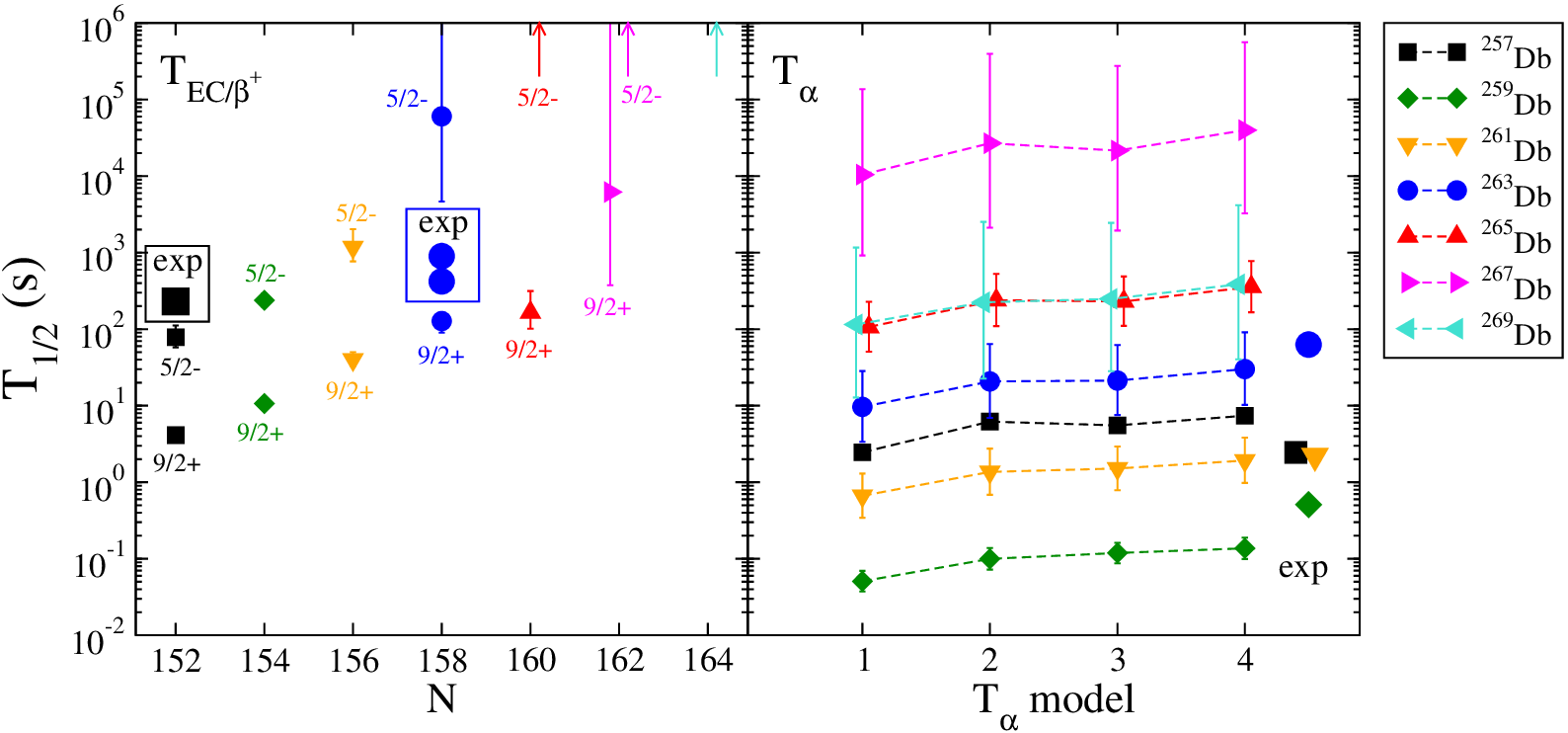}
\caption{Same as in figure \ref{fig_mdno}, but for the odd-$A$ Db isotopes 
from $^{257}$Db up to  $^{269}$Db.}
\label{fig_db}
\end{figure*}

%%%%%%%%%%%%%%%%%%%%%%%%%%%%Fig6%%%%%%%%%%%%%%%%%%%%%%%%%%%%%%%%%%%%%%%%%%%%%%%%%%%%%%%%
\begin{figure*}[htb]
\centering
\includegraphics[width=150mm]{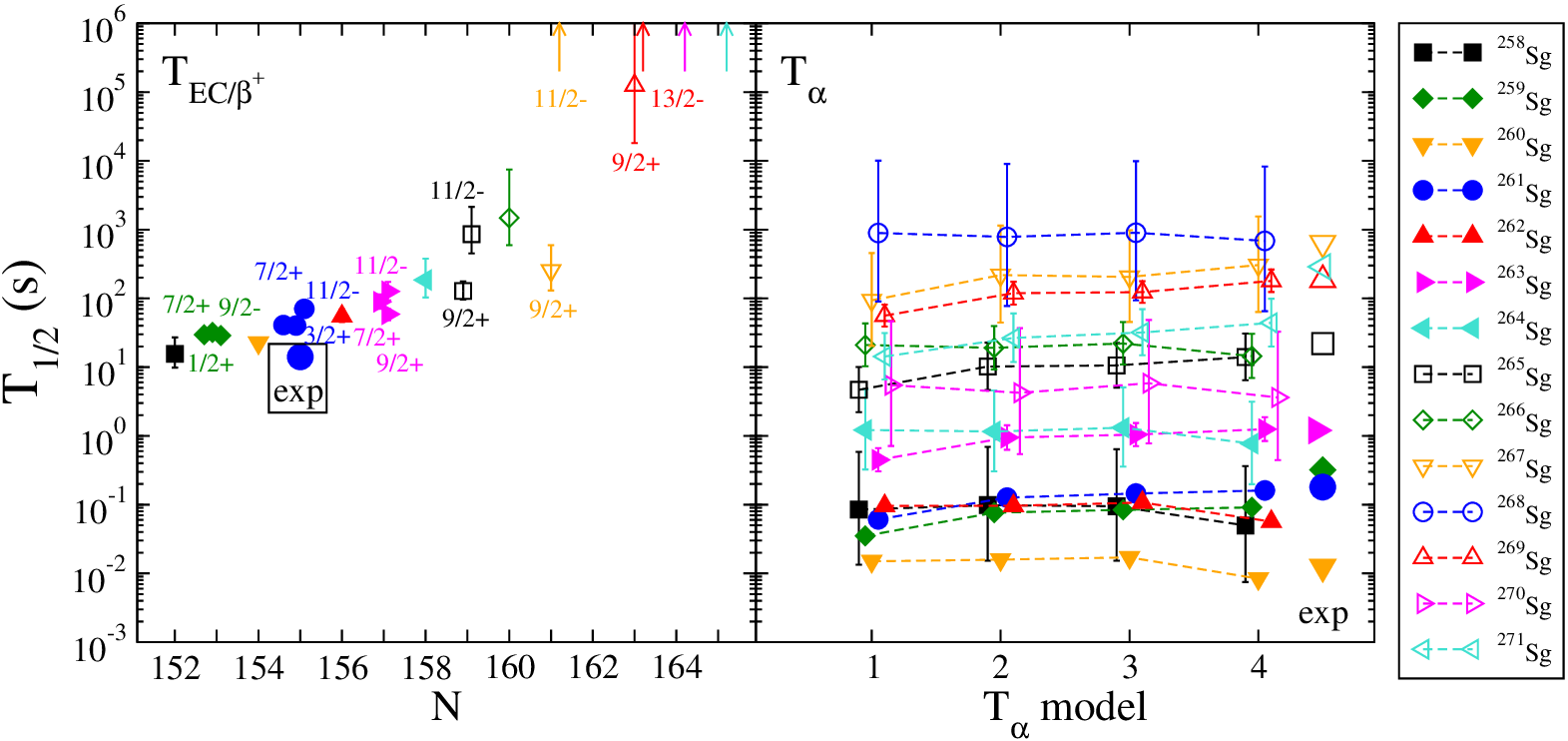}
\caption{Same as in figure \ref{fig_mdno}, but for $^{258-271}$Sg isotopes.}
\label{fig_sg}
\end{figure*}

%%%%%%%%%%%%%%%%%%%%%%%%%%%%Fig7%%%%%%%%%%%%%%%%%%%%%%%%%%%%%%%%%%%%%%%%%%%%%%%%%%%%%%%%
\begin{figure*}[htb]
\centering
\includegraphics[width=150mm]{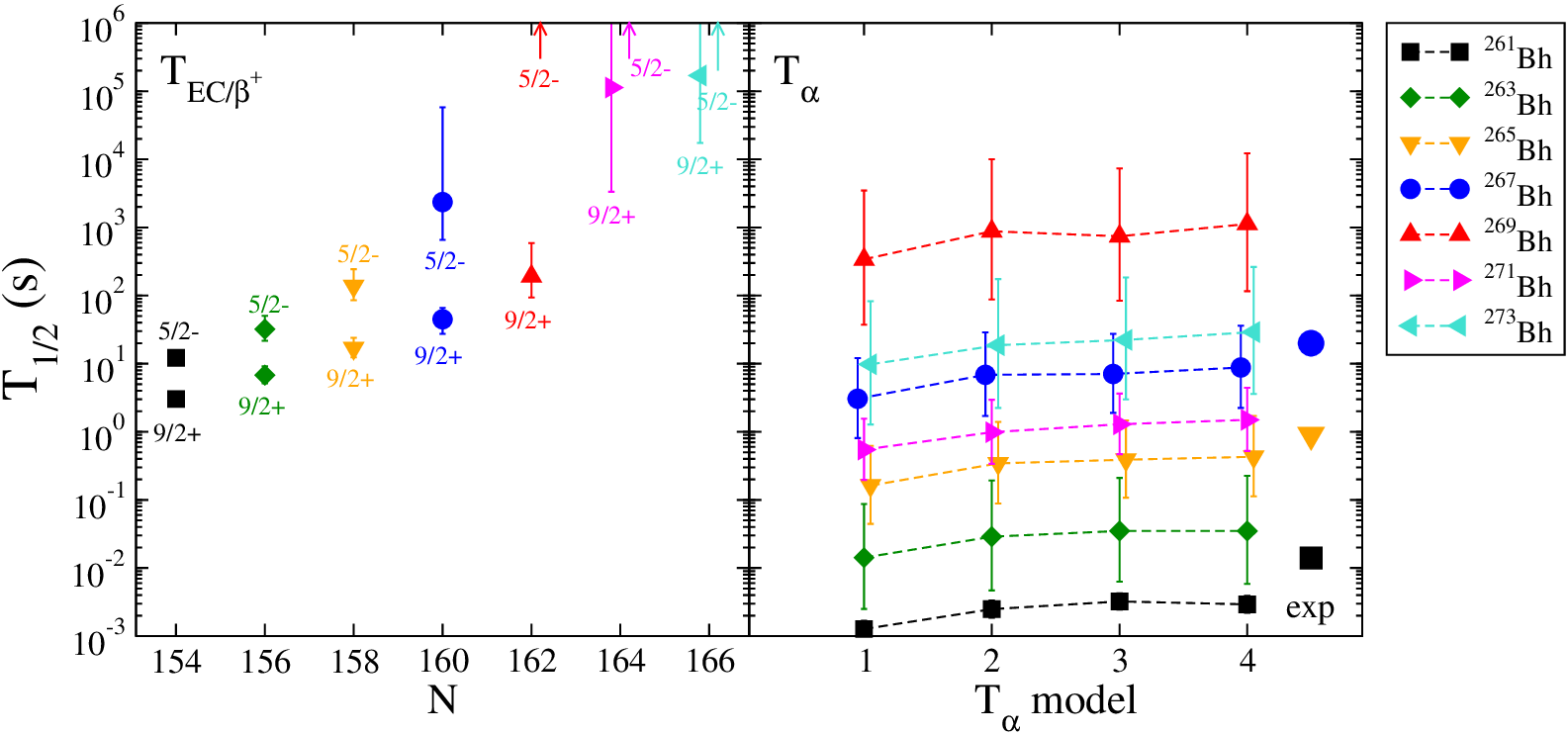}
\caption{Same as in figure \ref{fig_mdno}, but for the odd-$A$ Bh isotopes 
from $^{261}$Bh up to  $^{273}$Bh.}
\label{fig_bh}
\end{figure*}

%%%%%%%%%%%%%%%%%%%%%%%%%%%%Fig8%%%%%%%%%%%%%%%%%%%%%%%%%%%%%%%%%%%%%%%%%%%%%%%%%%%%%%%%
\begin{figure*}[htb]
\centering
\includegraphics[width=150mm]{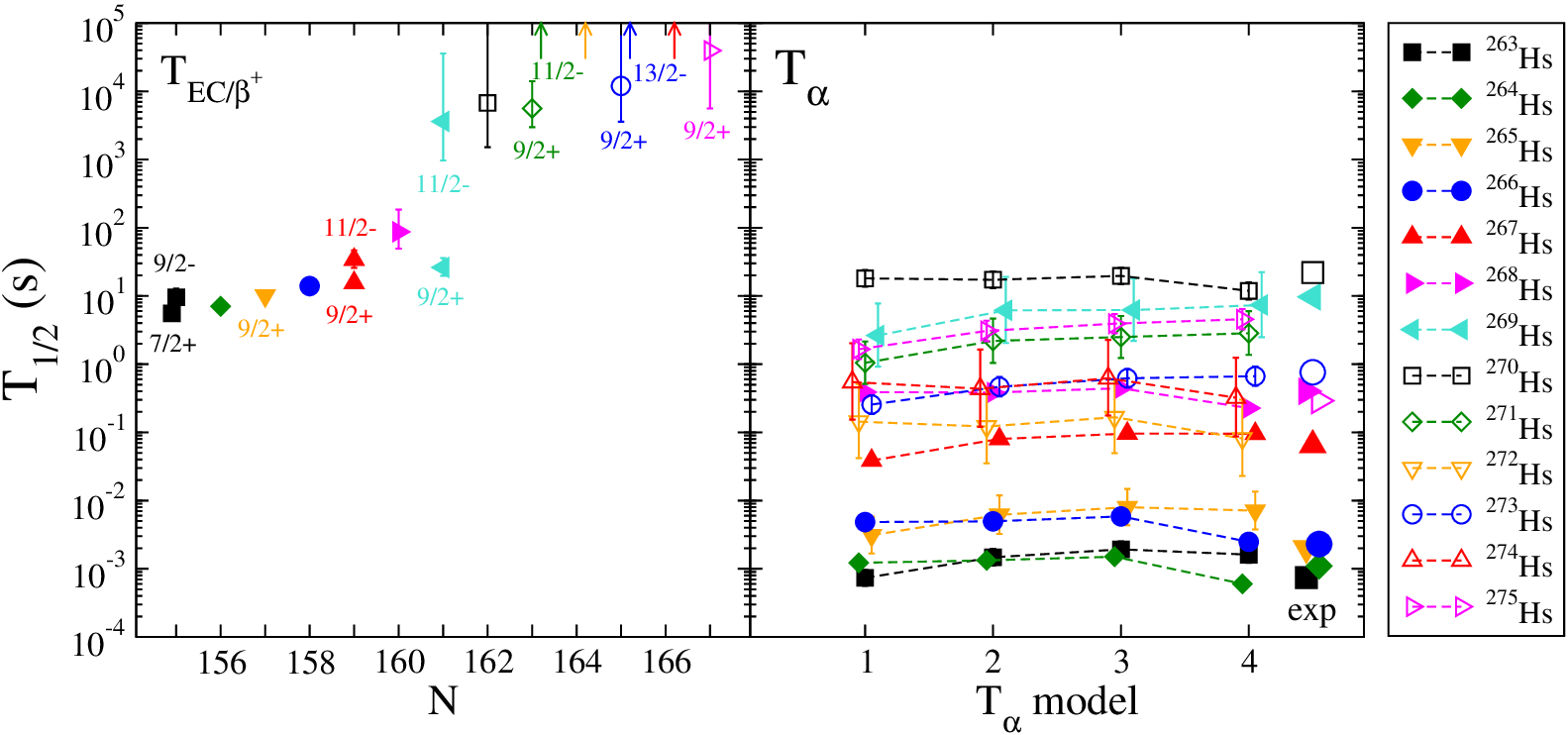}
\caption{Same as in figure \ref{fig_mdno}, but for $^{263-275}$Hs 
isotopes.}
\label{fig_hs}
\end{figure*}

%%%%%%%%%%%%%%%%%%%%%%%%%%%%Fig9%%%%%%%%%%%%%%%%%%%%%%%%%%%%%%%%%%%%%%%%%%%%%%%%%%%%%%%%
\begin{figure*}[htb]
\centering
\includegraphics[width=150mm]{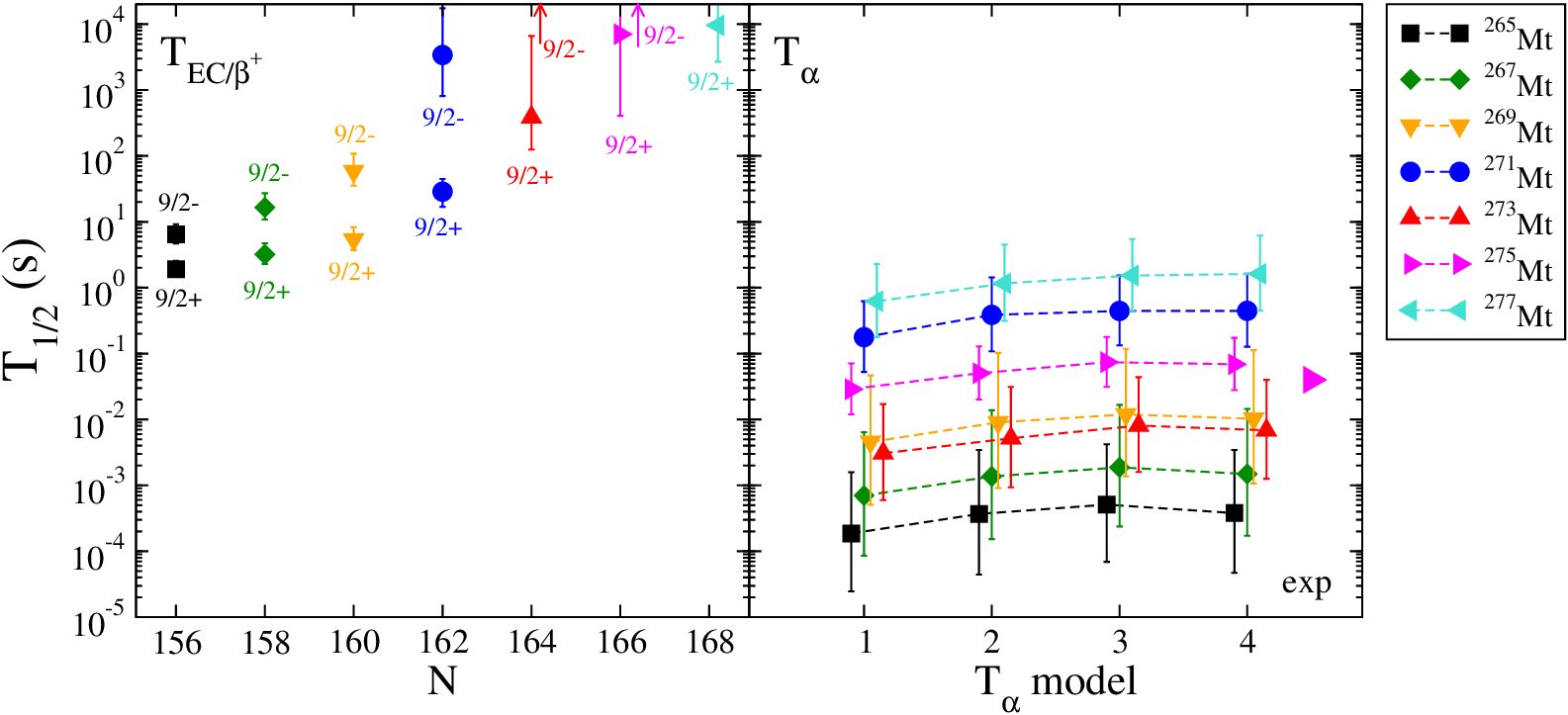}
\caption{Same as in figure \ref{fig_mdno},  but for the odd-$A$ Mt isotopes 
from $^{265}$Mt up to  $^{277}$Mt.}
\label{fig_mt}
\end{figure*}

%%%%%%%%%%%%%%%%%%%%%%%%%%%%Fig10%%%%%%%%%%%%%%%%%%%%%%%%%%%%%%%%%%%%%%%%%%%%%%%%%%%%%%%%

\end{document}